\documentclass[aps,prl,preprint,superscriptaddress,nopacs]{revtex4-1}
\usepackage[colorlinks, linkcolor=blue, anchorcolor=blue, citecolor=blue]{hyperref}
\usepackage{braket}
\usepackage{bm,color,amsmath,amssymb,mathrsfs,latexsym,graphicx,psfrag}

\newcommand{\bk}{{\mathbf k}}

\newcommand{\be}{\begin{equation}}
\newcommand{\ee}{\end{equation}}

\def\be{\begin{equation}}
\def\ee{\end{equation}}
\def\bea{\begin{eqnarray}}
\def\eea{\end{eqnarray}}

\begin{document}

\title{Weyl semimetal phase in non-centrosymmetric transition metal monophosphides  }

\author{Hongming Weng}
\email{hmweng@iphy.ac.cn}
\affiliation{Beijing National Laboratory for Condensed Matter Physics,
  and Institute of Physics, Chinese Academy of Sciences, Beijing
  100190, China}
\affiliation{Collaborative Innovation Center of Quantum Matter,
  Beijing, China}
  
\author{Chen Fang}
\affiliation{Department of Physics, Massachusetts Institute of Technology, Cambridge MA 02139, USA}

\author{Zhong Fang}
\affiliation{Beijing National Laboratory for Condensed Matter Physics,
  and Institute of Physics, Chinese Academy of Sciences, Beijing
  100190, China}
\affiliation{Collaborative Innovation Center of Quantum Matter,
  Beijing, China}
  
\author{Andrei Bernevig}
\affiliation{Department of Physics, Princeton University, Princeton NJ 08544, USA}

\author{Xi Dai}
\affiliation{Beijing National Laboratory for Condensed Matter Physics,
  and Institute of Physics, Chinese Academy of Sciences, Beijing
  100190, China}
\affiliation{Collaborative Innovation Center of Quantum Matter,
  Beijing, China}

\date{\today}

\begin{abstract}
{\bf 
Based on first principle calculations, we show that a family of nonmagnetic materials including TaAs, TaP, NbAs and NbP are Weyl semimetal (WSM) without inversion center. We find twelve pairs of Weyl points in the whole Brillouin zone (BZ) for each of them. In the absence of spin-orbit coupling (SOC), band inversions in mirror invariant planes lead to gapless nodal rings in the energy-momentum dispersion. The strong SOC in these materials then opens full gaps in the mirror planes, generating nonzero mirror Chern numbers and Weyl points off the mirror planes. The resulting surface state Fermi arc structures on both (001) and (100) surfaces are also obtained and show interesting shapes, pointing to fascinating playgrounds for future experimental studies. 
}
\end{abstract}


\maketitle

Most topological invariants
 in condensed matter non-interacting phases are defined on closed manifolds in momentum space. For gapped systems, both the Chern insulator and $Z_2$ topological insulator phases can be defined using the Berry phase and curvature in either the entire or half of the two-dimensional (2D) Brillouin zone (BZ) respectively.~\cite{TIreview, TIreview-2} 
  A similar idea can be generalized to gapless metallic systems. In three dimensional (3D) systems, besides the BZ, an important closed manifold in momentum space is a 2D Fermi surface (FS).  Topological metals can be defined by Chern numbers of the single-particle wave functions at the Fermi surface energies.~\cite{Na3Bi, Cd3As2, MRS:9383312} Such nonzero FS Chern number appears when the FS encloses a band crossing point $-$ Weyl point $-$ which can be viewed as a singular point of Berry curvature or ``magnetic monopole" in momentum space.~\cite{HgCrSe} 
Material with such Weyl points near the Fermi level are called Weyl semimetals(WSM).~\cite{ABJ_Weyl_1983, wan, Weyl_Physics}

 Weyl points can only appear when the spin doublet degeneracy of the bands is removed by breaking either time reversal $T$ or
 spacial inversion symmetry $P$ (in fact, Weyl points exist if the system does not respect $T\cdot P$). In these cases, the low energy single particle Hamiltonian around a
Weyl point can be written as a $2\times2$ ``Weyl equation", which is half of the Dirac equation in three dimension. According to the ``no-go Theorem",\cite{No_Go_I_1981,No_Go_II_1981}for any lattice model
the Weyl points always appear in pairs of opposite chirality or monopole charge. The conservation of chirality is one of the many ways to understand the topological stability of the WSM against any perturbation that preserve translational symmetry: the only way to annihilate a pair of Weyl points with opposite chirality is to move them to the same point in BZ. Since generically the Weyls can sit far away from each other in the BZ, this requires large changes of Hamiltonian parameters, and the WSM is stable.
The existence of Weyl points near the Fermi level will lead to several 
unique physical properties, including the appearance of discontinuous Fermi surfaces (Fermi arcs) on surface,~\cite{wan, Weyl_Physics,HgCrSe} 
the Adler-Bell-Jackiw anomaly~\cite{ABJ_Weyl_1983, PhysRevLett.111.246603, PhysRevLett.108.046602,PhysRevX.4.031035} and others.~\cite{Transport_Weyl_XLQi_2013, Weyl_node_line_super_Volovik}


The first proposal to realize WSM in condensed matter materials was suggested in Ref. ~\onlinecite{wan} for $Rn_2$Ir$_2$O$_7$ Pyrochlore with all-in/all-out magnetic structure, where twenty four pairs of Weyl points emerge as the system undergoes the magnetic ordering transition. A relatively simpler system HgCr$_2$Se$_4$~\cite{HgCrSe} has then been proposed by some of the present authors, where a pair of double-Weyl points with quadratic touching appear when the system is in a ferromagnetic phase. Another proposal involves a fine-tuned multilayer structure of normal insulator and magnetically doped topological insulators.~\cite{multilayerTRB} These proposed WSM systems involve magnetic materials, where the spin degeneracy of the bands is removed by breaking time reversal symmetry. As mentioned, the WSM can be also generated by breaking the spatial inversion symmetry only, a method which has the following advantages. First, compared with magnetic materials, nonmagnetic WSM are much more easily studied experimentally using angle resolved photo emission spectroscopy (ARPES) as alignment of magnetic domains is no longer required. Second, without the spin exchange field, the unique structure of Berry curvature leads to very unusual transport properties under strong magnetic field, unspoiled by the magnetism of the sample. 

Currently, there are three proposals for WSM generated by inversion symmetry breaking. One is a super-lattice system formed by alternatively stacking normal and topological insulators.~\cite{multilayerTRI} The second involves Tellurium or Selenium crystals under pressure~\cite{SeTe} and the third one is the solid solutions LaBi$_{1-x}$Sb$_x$Te$_3$ and LuBi$_{1-x}$Sb$_x$Te$_3$~\cite{LaBiSbTe3_Vanderbilt}  tuned around the topological transition points.~\cite{TI_NI_transition_Murakami_2007}
 In the present study, we predict that TaAs,TaP, NbAs and NbP single crystals are natural WSM and each of them possesses a total of 12 pairs of Weyl points.
 Compared with the existing proposals, this family of materials are completely stoichiometric, and therefore, are easier to grow and measure. 
 Unlike in the case of pyrochlore iridates and HgCr$_2$Se$_4$, where inversion is still a good symmetry and the appearance of Weyl points can be immediately inferred from the product of the parities at all the time reversal invariant momenta (TRIM),~\cite{Z2, inversion1, inversion2} 
in the TaAs family parity is no longer a good quantum number. However, the appearance of Weyl points can still be inferred by analyzing the mirror Chern numbers (MCN)~\cite{MCN_SnTe, MCN_YbB12} and $Z_2$ indices~\cite{Kane_z_2_2005, Z2} for the four mirror and time reversal invariant planes in the BZ. Similar with many other topological materials, the WSM phase in this family is also induced by a type of band inversion phenomena, which, in the absence of spin-orbit coupling (SOC), leads to nodal rings in the mirror plane. Once the SOC is turned on, each nodal ring will be gapped with the exception of three pairs of Weyl points leading to fascinating physical properties which include complicated Fermi arc structures on the surfaces.

\section{Crystal structure and band structure}

\begin{figure}[tbp]
\includegraphics[scale=0.4]{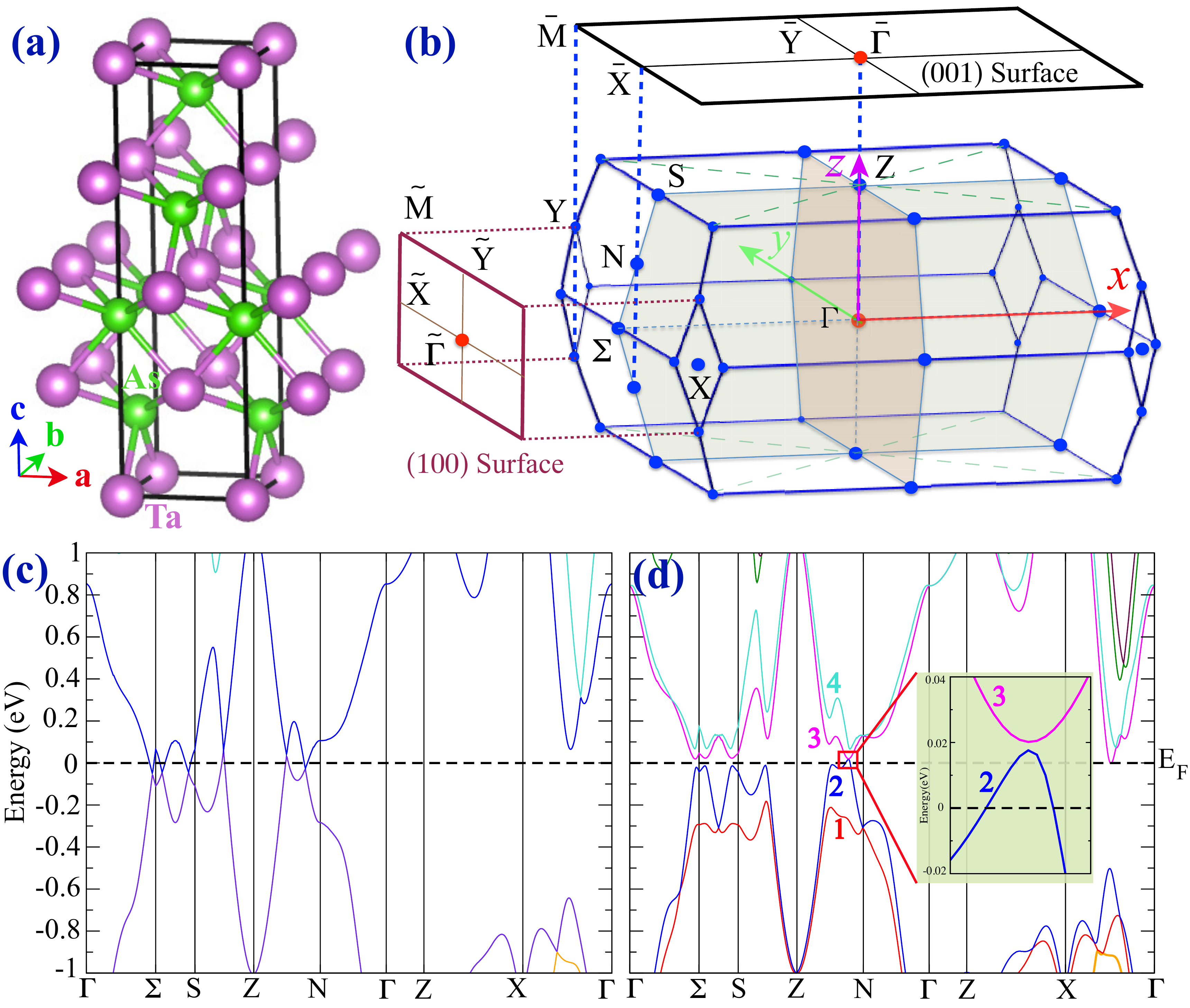}
\caption{{\bf Crystal structure and Brillouin zone (BZ)}. (a) The crystal symmetry of TaAs; (b) The bulk BZ and the projected surface BZ for both (001) and (100) surfaces; (c) The band structure of TaAs calculated by GGA without including the spin-orbit coupling; (d)The band structure of TaAs calculated by GGA with the spin-orbit coupling.}
\label{crystal_structure}
\end{figure}

As all four mentioned materials share very similar band structures, in the rest of the paper, we will choose TaAs as the representative material to introduce the electronic structures of the whole family.
The experimental crystal structure of TaAs~\cite{expCrystStru} is shown in Fig. ~\ref{crystal_structure}(a). 
It crystalizes in body-centered-tetragonal structure with nonsymmorphic space group 
$I4_{1}md$ (No.~109), which lacks inversion symmetry. The measured lattice constants are $a$=$b$=3.4348~\AA~and $c$=11.641 \AA. 
Both Ta and As are at $4a$ Wyckoff position (0, 0, $u$) with $u$=0 and 0.417 for Ta and As, respectively.
We have employed the software package OpenMX~\cite{openmx} for the first-principles calculation. It 
is based on norm-conserving pseudo-potential and pseudo-atomic localized basis functions. The choice of pseudopotentials, pseudo atomic orbital basis sets (Ta9.0-s2p2d2f1 
and As9.0-s2p2d1) and the sampling of BZ with $10\times10\times10$-grid have been carefully checked. The exchange-correlation functional within generalized 
gradient approximation (GGA) parameterized by Perdew, Burke, and Ernzerhof has been used.~\cite{PBE} After full structural relaxation, we obtain the lattice 
constants $a$=$b$=3.4824~\AA, $c$=11.8038 \AA~and optimized $u$=0.4176 for As site, in very good agreement with the experimental values. 
To calculate the topological invariant such as MCN and surface states of TaAs, we have generated atomic-like Wannier functions for Ta $5d$ and As $4p$ 
orbitals using the scheme described in Ref.~\onlinecite{myMLWF} 

We first obtain the band structure of TaAs without SOC by GGA and and plot it along the high symmetry directions in Fig. \ref{crystal_structure}(c). We  find clear band inversion and multiple band crossing features near the Fermi level along the ZN, ZS and $\Sigma$S lines. 
The space group of the TaAs family contains two mirror planes, namely $M_{x}$, $M_{y}$ (shaded planes in Fig.\ref{crystal_structure}(b) ) and two glide mirror planes,
namely $M_{xy}$, $M_{-xy}$ (illustrated by the dashed lines in Fig.\ref{crystal_structure}(b)).
The plane spanned by Z, N and $\Gamma$ points is invariant under mirror $M_y$ and the energy bands within the plane can be labeled by mirror eigenvalues $\pm 1$. 
Further symmetry analysis shows that the two bands that cross along the Z to N line belong to opposite mirror eigenvalues and hence the crossing between them is protected by mirror symmetry. Similar band crossings can also be found along other high symmetry lines in the  
ZN$\Gamma$ plane, i.e. the ZS and NS lines. Altogether these band crossing points form  a ``nodal ring" in the ZN$\Gamma$ plane as shown in Fig.\ref{band_structure}(b). 
Unlike for the situation in ZN$\Gamma$ plane, in the two glide mirror planes ($M_{xy}$ and $M_{-xy}$), the band structure is fully gaped with the minimum gap of roughly 0.5 eV.


The analysis of orbital character shows that the bands near the Fermi energy are mainly formed by Ta $5d$ orbitals, which have large SOC. Including SOC in the first principle calculation leads to a dramatic change of the band structure near Fermi level, as plotted in Fig.~\ref{crystal_structure}(d). At first glance, it seems that the previous band crossings in the ZN$\Gamma$ plane are all gaped with the exception of one point along ZN line. Detailed symmetry analysis reveals that the bands ``2" and ``3" in Fig.\ref{crystal_structure}(d) belong to opposite mirror eigenvalues, 
indicating the almost touching point along the ZN line is completely accidental. In fact there is a small gap of roughly  3 meV between bands ``2" and ``3" as illustrated by the inset of Fig.\ref{crystal_structure}(d). The ZN$\Gamma$ plane then becomes fully gapped once SOC is turned on.

\section{Topological invariants for mirror plane and Weyl points}

\begin{figure}[tbp]
\includegraphics[width=0.75\textwidth]{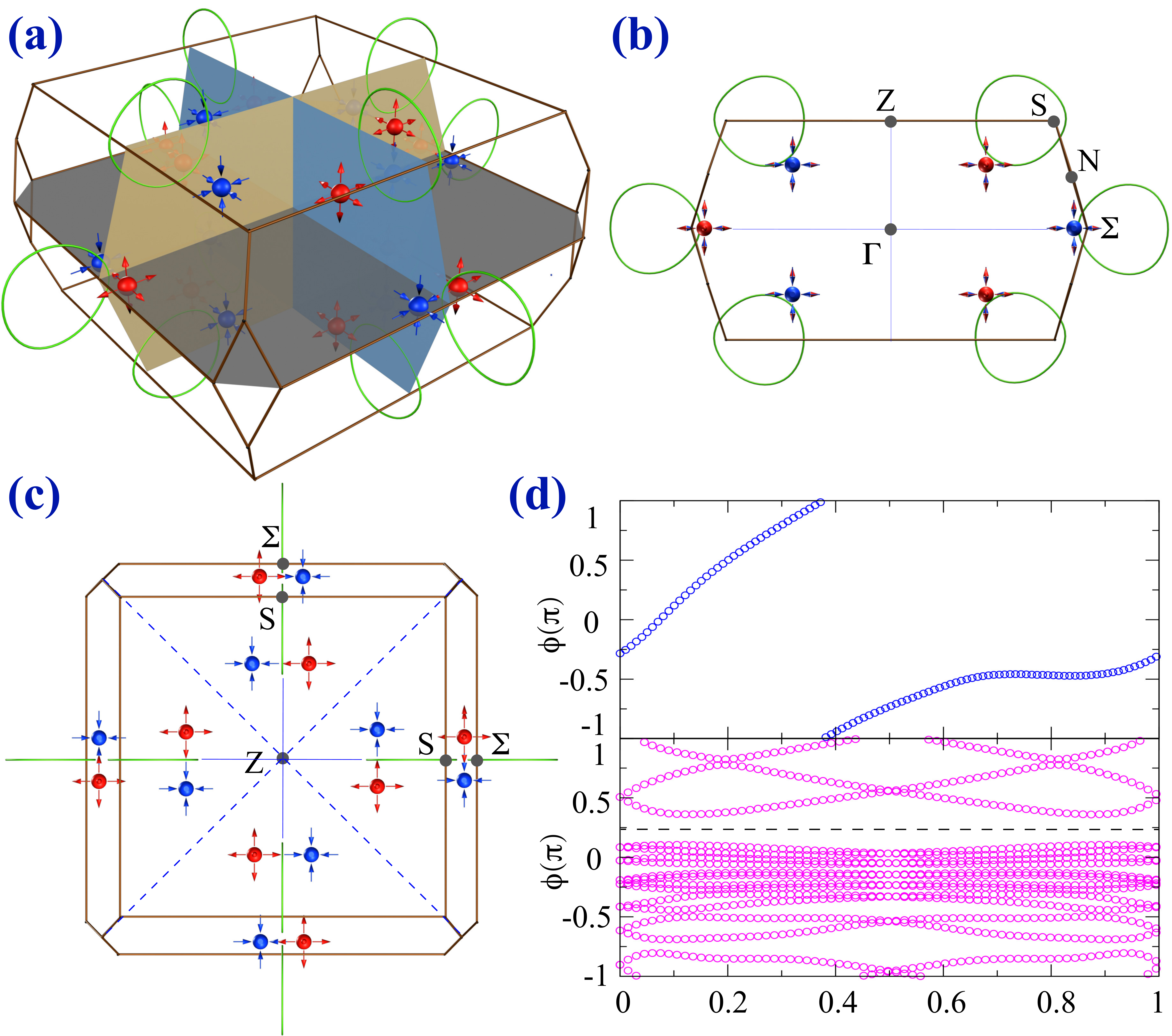}
\caption{{\bf Nodal rings and Weyl points distribution, as well as $Z_2$ and MCN for mirror planes}. (a) 3D view of the nodal rings (in the absence of SOC) and Weyl points (with SOC) in the BZ; (b) Side view from [100] and (c) top view from [001] directions for the nodal rings and Weyl points. 
Once the SOC is turned on, the nodal rings are gapped and give rise to Weyl points off the mirror planes (see movie in Appendix); 
(d) top panel: Flow chart of the average position of the Wannier centers obtained by Wilson loop calculation for bands 
with mirror eigenvalue $i$ in the mirror plane ZN$\Gamma$; bottom panel: The flow
chart of the Wannier centers obtained by Wilson loop calculation for bands in the glide mirror plane ZX$\Gamma$. 
There is no crossing along the reference line (the dashed line) indicating the $Z_2$ index is even.
}
\label{band_structure}
\end{figure}

Since the material has no inversion center, the usual parity condition ~\cite{Z2, inversion1, inversion2} can not be applied to predict the existence of WSM. We then resort to another strategy. As previously mentioned, the space group of the material provides two mirror planes ($M_x$ and $M_y$), 
where the MCN can be defined. If a full gap exists for the entire BZ, the MCN would directly reveal whether this system is a topological crystalline insulator or not. Interestingly, as shown below, if the system is not fully gaped we can still use the MCN to find out whether the material hosts Weyl points in the BZ or not. Besides the two mirror planes, we have two additional glide mirror plane ($M_{xy}$ and $M_{-xy}$).
 Although the MCN is not well defined for the glide mirror planes, the $Z_2$ index is still well defined here 
 as these planes are time reversal invariant. We then apply the Wilson loop method to calculate the MCNs
 for the two mirror planes and $Z_2$ indices for the two glide mirror planes.
For a detailed explanation of the method, please refer to Ref.~\onlinecite{YuRui_Z2_2011PRB, MRS:9383312}. The results are plotted in Fig.\ref{band_structure}(d), which shows that MCN is one for ZN$\Gamma$ plane ($M_y$) and $Z_2$ index is 
even or trivial for ZX$\Gamma$ plane($M_{xy}$). Then, if we consider the $(001)$ surface, which is invariant under the $M_y$ mirror.
The non-trivial helical surface modes will appear due to the non-zero MCN in  ZN$\Gamma$ plane, which generates a single pair of
FS cuts along the projective line of the ZN$\Gamma$ plane (the $x$-axis in Fig.\ref{band_structure}(c)).
Whether these Fermi cuts will eventually form a single closed Fermi circle or not 
depends on the $Z_2$ index for the two glide mirror plane, which are projected to the dashed blue lines in
Fig.\ref{band_structure}(c). Since the $Z_2$ indices for the glide mirror planes are trivial, as confirmed by our Wilson loop 
calculation plotted in Fig.\ref{band_structure}(d), there are no protected helical edge modes along the projective lines of the 
glide mirror planes  (dashed blue lines in Fig.\ref{band_structure}(c)) and the Fermi cuts along the $x$-axis in 
Fig.\ref{band_structure}(c) must end somewhere between the $x$-axis and the diagonal lines (dashed blue lines in Fig.\ref{band_structure}(c)). 
In other words they must be Fermi arcs,
indicating the existence of Weyl points in the bulk band structure of TaAs.

\begin{figure}[tbp]
\includegraphics[width=0.40\textwidth]{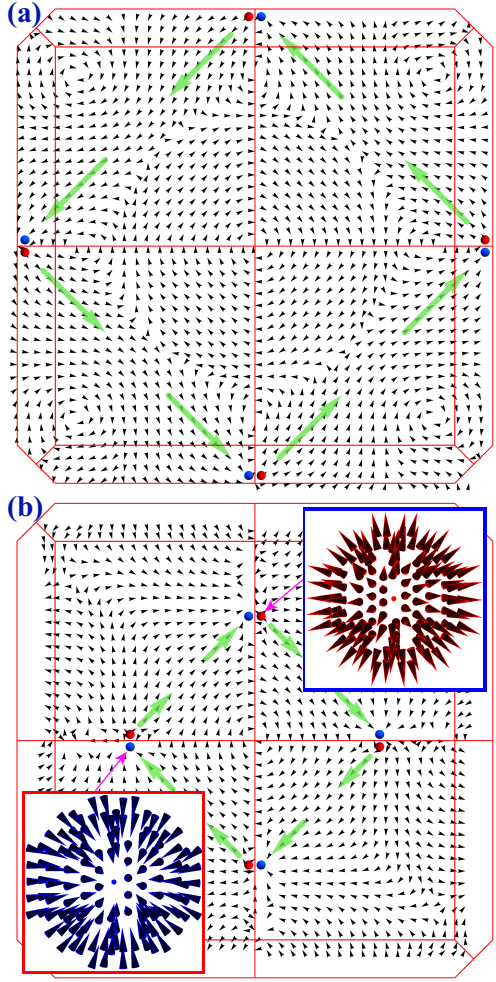}
\caption{{\bf Berry curvature from pairs of Weyl points}. (a) The distribution of the Berry curvature for the $k_z=0$ plane, where the blue and red dots denote the Weyl points with chirality of +1 and -1, respectively; (b) same with (a) but for  $k_z= 0.592\pi$ plane. The insets show the 3D view of hedgehog like Berry curvature near the two selected Weyl points.}
\label{Berry_Curvature}
\end{figure}

From the above analysis of the MCN and $Z_2$ index of several high-symmetry planes, we can conclude that Weyl points exist in the TaAs band structure. We now determine the total number of the Weyls and their exact positions. This is a hard task, as the Weyl points are located at generic $k$ points without any little-group symmetry. For this purpose, we calculate the integral of the Berry curvature on closed surface in k-space, which equals the total chirality of the Weyl points enclosed by the given surface. 
Due to the four fold rotational symmetry and mirror planes that characterize TaAs, 
we only need to search for the Weyl points within the reduced BZ - one eighth of the whole BZ. We first calculate the
total chirality or monopole charge enclosed in the reduced BZ. The result is one, which guarantees the existence of, and odd number of Weyl points.
 To determine precisely the location of each Weyl point, we divide the reduced BZ into a very dense $k$-point mesh and compute 
 the Berry curvature or 
the ``magnetic field in momentum space"~\cite{BC_Wannier, myMLWF} on that mesh as shown in Fig.\ref{Berry_Curvature}. From this, we can easily identify the precise position of the Weyl points by searching for
 the ``source" and ``drain" points of the ``magnetic field". The Weyl points in TaAs are illustrated in Fig.\ref{band_structure}(a), where we find 12 pairs of Weyl points
  in the vicinity of what used to be, in the SOC-free case, the nodal rings on two of the mirror invariant planes. For each of the mirror invariant planes, after turning on SOC,
 the nodal rings will be fully gaped within the plane but isolated gapless nodes slightly off plane appear as illustrated in  Fig.\ref{band_structure}(b). Two pair of Weyl points are located exactly in the $k_z=0$ plane and another four pairs of Weyl points are located off the $k_z=0$ plane.
 Considering the four-fold rotational symmetry, it is then easy to understand that there are totally 12 pairs of Weyl points in the whole BZ. 
 The appearance of Weyl points can also be derived from a $k\cdot p$ model with different types of mass terms induced by SOC, which will be introduced in detail in the Appendix. 
 The band structures for the other  three materials TaP, NbAs and NbP are very similar. The precise positions of the Weyl points for all these materials are summarized in Table.~I. 
 
 \begin{table}
 \label{weylpoint}
\caption{The two nonequivalent Weyl points in the $xyz$ coordinates shown in Fig. 1(b). The position is given in unit of the length of $\Gamma$-$\Sigma$ for $x$ and $y$ and of the length of $\Gamma$-Z for $z$.}\label{WeylPosition}
\begin{tabular*}{0.45\textwidth}{@{\extracolsep{\fill}}c|c|c}
 	\hline\hline
    & Weyl Node 1 & Weyl Node 2  \\
    \hline
TaAs & (0.949, 0.014, 0.0) & (0.520, 0.037, 0.592) \\
 	\hline
TaP & (0.955, 0.025, 0.0) & (0.499, 0.045, 0.578) \\
 	\hline
NbAs & (0.894, 0.007, 0.0) & (0.510, 0.011, 0.593) \\
 	\hline
NbP  & (0.914, 0.006, 0.0) & (0.494, 0.010, 0.579) \\
 	\hline\hline
\end{tabular*}
\end{table}

\section{Fermi arcs and surface states}

Unique surface states with unconnected Fermi arcs can be found on the surface of a WSM. These can be understood in the following way:
for any surface of a WSM, we can consider small cylinders in the momentum space parallel to the surface normal. In the 3D BZ, these cylinders will be cut by the zone boundary and their topology is equivalent to that of a closed torus rather than that of open cylinders. If a cylinder encloses a Weyl point, by Stokes theorem, the total integral of the Berry curvature (Chern number) of this closed torus must equal the total ``monopole charge" carried 
by the Weyl point(s) enclosed inside. On the surface of the material, such a cylinder will be projected to a cycle surrounding the projection point
of the Weyl point and a single Fermi surface cut stemming from the chiral edge model of the 2D manifold with Chern number 1 (or -1) must be found on that circle.
By varying the radius of the cylinder, it is easy to show that such FSs must start and end at the projection of two (or more) Weyl points with different ``monopole
charge" , i.e. they must be ``Fermi arcs".~\cite{wan, HgCrSe, Transport_Weyl_XLQi_2013}
In the TaAs materials family, on most of the common surfaces, multiple Weyl points will be projected on top
of each other and we must generalize the above argument to multiple projection of Weyl points. 
It is easy to prove that the total number of surface modes at Fermi level crossing 
a closed circle in surface BZ must equal to the sum of 
 the ``monopole charge" of the Weyl points inside the 3D cylinder that projects to the given circle.
 Another fact controlling the behavior of the surface states is the MCN introduced in the previous discussion, which limits the number of FSs cutting certain 
projection lines of the mirror plane (when the corresponding mirror symmetries are still preserved on the surface).

\begin{figure}[tbp]
\includegraphics[width=0.75\textwidth]{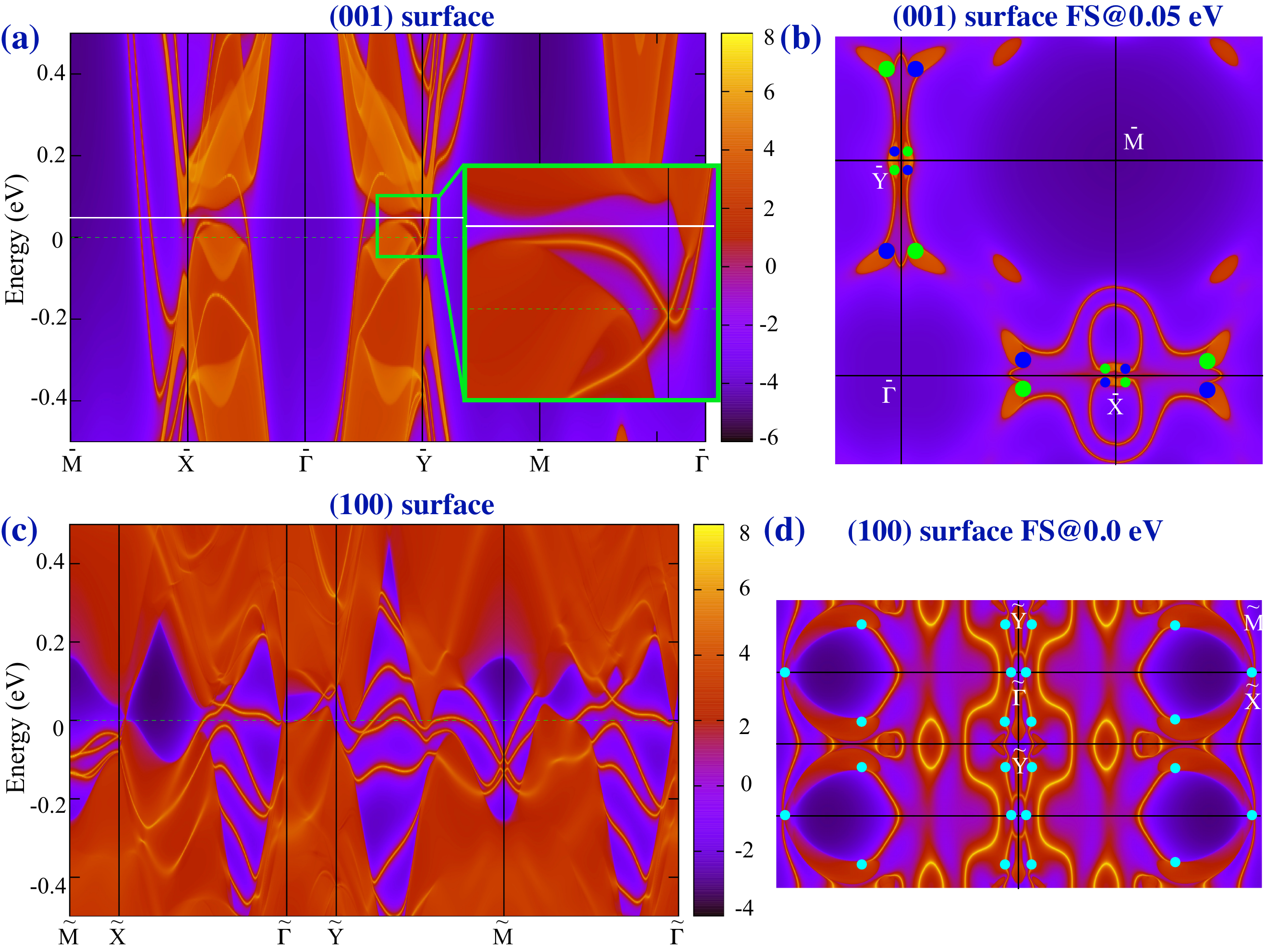}
\caption{{\bf Fermi arcs in the surface states}. (a) The Surface states for (001) surface; (b) the corresponding fermi surfaces on (001) surfaces; 
(c) The Surface states for (100) surface. The dots illustrate the 
projective points of the bulk Weyl points on the surfaces, where the color represents the 
chirality of the Weyl points (blue for positive and green for negative), 
small dot represents the single projected Wely point, and
large dot represents two Weyl points with same chirality projecting
on top of each other.
(d) the corresponding fermi surfaces on (100) surfaces. All the dots here are the projective points for a pair of Weyl points with opposite chirality.  }
\label{surf_state}
\end{figure}

By using the Green's function method~\cite{MRS:9383312} based on the tight-binding (TB) Hamiltonian generated by the previously obtained Wannier functions, we have computed the surface states for both (001) and (100) surfaces. They are plotted in Fig.~\ref{surf_state} together with the FS plots. On the (001) surface, the crystal symmetry
is reduced to $C_{2v}$ leading to different behavior for the surface bands around $\bar X$ and $\bar Y$ points respectively. 
Along the $\bar\Gamma$-$\bar X$ or $\bar\Gamma$-$\bar Y$ lines, 
there are two FS cuts with
the opposite Fermi velocity satisfying the constraint from the MCN for $\Gamma$ZN plane. In addition to the MCN, the possible ``connectivity pattern" of the Fermi arcs on the
surface has to link different projection points of the Weyl nodes in a way that obeys the chirality condition discussed in the above paragraph. 
For the (001) surface of TaAs, the connectivity pattern of the Fermi arcs that satisfy all the conditions discussed is not unique.  However, due to the fact that
all the projective points on the (001) surface are generated either by a single Weyl
point or by two Weyl points with the same chirality, the appearance of ``Fermi arcs" on the (001) surface is guaranteed.
The actual Fermi arc connectivity pattern for (001) surface is shown in Fig.~\ref{surf_state}(b), obtained by our $ab$-$initio$ calculation on a non
relaxed surface described by the TB model. Changes of surface potentials or the simple relaxation of the surface charge density might lead to transitions of those Fermi arc 
connectivity pattern and result in topological Fermi arc phase transitions on the surface. A very interesting point of the (001) surface states  is the extremely long Fermi arcs which cross the zone boundary along the $\bar X$ to $\bar M$ line. Comparing with other proposed WSM materials, the Fermi arcs in TaAs families are much longer, which greatly facilitates their detection in experiments. 

Comparing to the (001) surface, the surface states in (100) surface of TaAs are much more complicated as shown in Fig.~\ref{surf_state}(c) and (d).
The biggest difference between the (100) and (001) surfaces is that all the projected Weyl points on the (100) surface are 
formed by a pair of Weyl points with opposite chirality, which does not guarantee (but does not disallow) the existence of the Fermi arcs. The only
constraint for the (100) surface states is the nonzero MCN of the $\Gamma ZN$ plane, which generates a pair of chiral modes along the
$\bar\Gamma \bar Y$ line, the projection of the mirror plane, as illustrated in Fig.\ref{surf_state}(d).

\section{Conclusion}

In summary, a family of nonmagnetic WSM materials are proposed in the present paper. Each material in this family contains 
12 pairs of Weyl points, which appear due to the lack of inversion center in the crystal structure and can be derived from the nonzero MCN for
two of the mirror invariant planes in the BZ. The surface states of these materials 
form quite complicated patterns for the Fermi surfaces, which are
determined by the both the chirality distribution of the Weyl points and the MCNs for the mirror invariant planes.



\section{acknowledgments}
We acknowledge helpful discussions with A. Alexandradinata and N. P. Ong, and the help in plotting Fig.~\ref{Berry_Curvature} from Rui Yu.
This work was supported by the National Natural Science Foundation of China, the 973 program of China (No. 2011CBA00108 and No.
2013CB921700), and the ``Strategic Priority Research Program (B)" of the Chinese Academy of Sciences (No. XDB07020100). 
The calculations were preformed on TianHe-1(A), the National Supercomputer Center in Tianjin, China. BAB acknowledges 
support from NSF CA-REER DMR-0952428, ONR-N00014-11-1-0635, MURI-130-6082, DARPA under SPAWAR Grant No.: N66001-11-1-4110, the Packard Foundation.

\newpage
\clearpage
\section{Appendix}
\subsection{I. $k\cdot{p}$ Model of a Nodal Ring and the Appearance of Weyl Points Due to Spin-Orbit Coupling}
We perform a $k\cdot{p}$ analysis in the vicinity of the $\Sigma$-point in the 3D BZ. We show that (i) a nodal ring (a closed line of band-touching points) is protected in the presence of a mirror reflection symmetry and SU(2) spin rotation symmetry (when spin-orbit coupling is absent)  and (ii) when spin-orbital coupling (SOC) is turned on and all crystalline symmetries are preserved, the nodal ring may be partially gapped with several distinct possibilities: (i) Weyl nodes on the $(001)$-plane, (ii) Weyl nodes away from the $(001)$-plane and (iii) nodal rings on the $(010)$-plane and (iv) full gap. The realization of these different possibilities strongly depends on the specific form of SOC terms.

The nodal ring around $\Sigma$-point can be modeled by a two-band $k\cdot{p}$-theory, in the absence of SOC:
\bea\label{eq:kdotp}
H_0(\bk)=\sum_{i=x,y,z}d_i(\bk)\sigma_i,
\eea
where $d_i(\bk)$ are real functions, and $\bk=(k_x,k_y,k_z)$ are three components of the momentum $\bk$ relative to $\Sigma$-point along $[100]$-, $[010]$- and $[001]$-axes, respectively. In Eq.(\ref{eq:kdotp}) we have ignored the kinetic term proportional to the identity matrix, as it is irrelevant in studying the band-touching. The mirror reflection symmetry, denoted by $M_{010}$, is represented by $M=\sigma_z$: in the absence of SOC, we can choose a mirror symmetry that squares to unity. The form of the mirror operator is chosen such that the two bands have opposite mirror eigenvalues, an information obtained from the ab-initio calculation. The mirror reflection dictates that
\bea
MH_0(k_x,k_y,k_z)M^{-1}=H_0(k_x,-k_y,k_z),
\eea
which translates into
\bea\label{eq:dconstraints}
d_{x,y}(k_x,k_y,k_z)&=&-d_{x,y}(k_x,-k_y,k_z),\\
d_z(k_x,k_y,k_z)&=&d_z(k_x,-k_y,k_z).
\eea
Eq.(\ref{eq:dconstraints}) states that on the plane $k_y=0$, only $d_z$ is nonzero, and hence generically the equation $d_z(k_x,0,k_z)=0$ will have codimension of one, i.e., nodal line solution. Symmetry preserving perturbations involve gradually changing the forms of $d_i$'s without violating Eq.(\ref{eq:dconstraints}), so the nodal ring is robust against them.

Another symmetry is present at the $\Sigma$-point: a twofold rotation $C_2$ about $[001]$-axis followed by time-reversal symmetry. This is because the rotation sends the $\Sigma$-point to its time-reversal partner, and a further time-reversal sends it back. This symmetry may be represented by $C_{2T}=U_TK$, where $U_T$ is any symmetric and unitary matrix, and $K$ is complex conjugation. Without SOC, the rotation about $[001]$-axis and the reflection about $(010)$-plane commute with each other, so we require
\bea
[\sigma_z,KU_T]=0,
\eea
and we may choose $U_T=\sigma_z$. This symmetry places additional constraints on $d_i$'s:
\bea
H_0(k_x,k_y,k_z)=\sigma_zH^\ast_0(k_x,k_y,-k_z)\sigma_z,
\eea
or
\bea\label{eq:dconstraints2}
d_{y,z}(k_x,k_y,k_z)=d_{y,z}(k_x,k_y,-k_z),\\
\nonumber
d_x(k_x,k_y,k_z)=-d_x(k_x,k_y,-k_z).
\eea
Eqs.(\ref{eq:dconstraints},\ref{eq:dconstraints2}) determine the general form of our $k\cdot{p}$-model.

Now we consider adding spin-orbit coupling terms while respecting the symmetries at $\Sigma$-point. We first need to determine the matrix representations of the generators of the little group, i.e., $M_{010}$ and $C_2*T$. Considering spin degrees of freedom, we know that (i) a mirror reflection consists of a spatial reflection and a twofold spin rotation about the axis perpendicular to the reflection plane, (ii) a twofold rotation involves a spatial twofold rotation and a twofold spin rotation about the same axis and (iii) time-reversal symmetry involves complex conjugation and a flipping of the spin. Following these facts we obtain the matrix representations:
\bea
M&=&i\sigma_z\otimes{s}_y,\\
\nonumber
C_{2T}&=&K\sigma_z\otimes{s}_x.
\eea Notice that now $M^2=-1$ and $C_{2T}^2=1$, as needed. 
With spin degrees of freedom, each band in the previous spin-orbit coupling free model in Eq.[\ref{eq:kdotp}] becomes two bands, and nodal ring becomes a four-band crossing. In the vicinity of the nodal ring, the addition of SOC is equivalent to adding coupling between different spin components, i.e., `mass terms', to the previous model. Here the name `mass term' simply means that these terms are not required to vanish at the nodal ring by any symmetry.

The symmetry of the nodal ring is just mirror reflection. The mass terms hence must commute with mirror symmetry, and a generic term on the $k_y=0$ plane is given by
\begin{widetext}
\bea
H_m=m_1(\mathbf{k})s_y+m_2(\mathbf{k})\sigma_zs_y+m_3(\bk)\sigma_xs_x+m_4(\bk)\sigma_xs_z+m_5(\bk)\sigma_ys_z+m_6(\mathbf{k})\sigma_ys_x.
\eea
\end{widetext}
Note that these mass terms are in general $\bk$-dependent, as their values may change as $\bk$ moves along the nodal ring, but the $C_2*T$ symmetry makes them satisfy (on the $k_y=0$ plane)
\begin{widetext}
\bea\label{eq:msymm}
m_{1,2,4,6}(k_x,0,k_z)&=&m_{1,2,4,6}(k_x,0,-k_z)=m_{1,2,4,6}(k_x,0,k_z),\\
\nonumber
m_{3,5}(k_x,0,k_z)&=&-m_{3,5}(k_x,0,-k_z)=m_{3,5}(k_x,0,k_z).
\eea
\end{widetext} 
A complete analysis of the band crossing in the presence of all six mass terms is unavailable as the analytic expressions for the dispersion are involved. However, one may see the qualitative role played by each mass term by analyzing them separately. From Eq.(\ref{eq:msymm}), we see that $m_{3,5}$ are odd under $k_z\rightarrow-k_z$, while the others are even. This indicates that only $m_{1,2,4,6}$-terms are responsible for band crossings appearing on the $k_z=0$ plane, while the band crossings away from that plane are attributed mainly to the presence of $m_{3,5}$-terms.

At $k_y=0$ plane, $m_{1,2}$-terms commute with $H_0$, so these terms, if of small strength, will split the doubly degenerate nodal ring into two singly degenerate rings, but not open gaps. The equations for the two new rings are given by
\bea
d_{z}(k_x,0,k_z)\pm{m_{1,2}}=0.
\eea
One should note that when $m_{1}$-term ($m_2$-term) is added, the two rings are the crossing between two bands with same (opposite) mirror eigenvalues. Therefore, the two rings from adding $m_1$ term are purely accidental, and the two rings from adding $m_2$ term are protected by mirror symmetry.

Next we discuss the effect of $m_{4,6}$ terms, which should in combination gives rise to the pair of Weyl nodes on the $k_z=0$-plane shown in the paper. The dispersion after adding $m_{4,6}$ terms is
\bea
E(\bk)=\sqrt{d^2+m_4^2+m_6^2\pm2\sqrt{m_4^2d_x^2+m_6^2d_y^2+m_4^2m_6^2}},
\eea
where $d^2=d_x^2+d_y^2+d_z^2$. With some straightforward algebraic work, it can be shown that the equation $E(\bk)=0$ (band-touching) is equivalent to the following three equations:
\bea
d_x&=&d_z=0,\\
\nonumber
d_y^2&=&m_6^2-m_4^2.
\eea
When $|m_6|>|m_4|$, these equations have at least one pair of solutions on $k_z=0$-plane symmetric about $k_y=0$ with codimension zero: they are Weyl nodes on the $k_z=0$ plane. In our simulation, we found only one pair of Weyl nodes appear on this plane, which can only be understood if $m_{4,6}$ are $\bk$-dependent. The equations $d_x=0$ and $d_z=0$ determines a closed loop on the $k_z=0$ plane. At the same time $d_y(k_x,k_y,0)=\pm\sqrt{m_6^2-m_4^2}$ has solutions that are symmetric about $k_y=0$. Since $d_y(k_x,0,0)=0$, the solutions do not cross the $k_y=0$ line if $m_{4,6}$ are constants. Therefore, the solutions must be two lines, which make four crossings in total with the solution to $d_z=0$. However, let us recall that all mass terms can also contain a linear function in $k_x$, so it is possible that $m_6-m_4$ vanishes for a particular $k_x$. At that $k_c$, the solution to $d_y=\sqrt{m_6^2-m_4^2}=0$ is satisfied at $k_y=0$. If the point $(k_c,0,0)$ is inside the loop that solves $d_z(k_x,k_y,0)=0$, then there must be an odd number of crossings of $d_y=\sqrt{m_6^2-m_4^2}$ and $d_z=0$.

We can also understand the pairs of Weyl nodes that are away from $k_z=0$-plane. Consider a coexistence of both $m_4$- and $m_5$-terms. The dispersion is given by
\bea
E(\bk)=\sqrt{(d_x\pm{m_4})^2+(d_y\pm{m_5})^2+d_z^2}.
\eea
Solving $E(\bk)=0$ is equivalent to solving
\bea
d_z&=&0,\\
\nonumber
d_x&=&\pm{m_4},\\
\nonumber
d_y&=&\pm{m_5}.
\eea
The last two equations together give
\bea
k_z^2=\sqrt{vm_4/\lambda},\\
k_y^2=\sqrt{m_4\lambda/v},
\eea
where $d_x\equiv{u}{k}_yk_z$, $m_5\equiv\lambda{k_z}$ and $d_y\equiv{v}k_y$.

When $vm_4\lambda<0$, there is no solution. When $vm_4\lambda<0$, there are four sets of solutions for $(k_y,k_z)$. We can substitute them into $d_z=0$ to obtain the four Weyl nodes observed in our simulation. When $m_4$ is small, the four Weyl nodes are close to the crossing point of the nodal ring and the $k_z=0$ plane.

We summarize the roles played by different mass terms: $m_{1,2}$-terms split the nodal ring into two non-degenerate rings. With $m_1$- ($m_2$-)term alone the ring is the crossing of two bands with same (opposite) mirror eigenvalues. $m_{3,5}$-terms gap the nodal ring except at $k_y=k_z=0$. $m_4$-term alone or coexisting with $m_3$-term fully gaps the ring. $m_4$-term coexisting with $m_5$ produces four Weyl nodes away from the $k_z=0$ plane. $m_6$-term creates a pair of Weyl nodes on the $k_z=0$ plane, symmetric about $k_y=0$.

\subsection{II. Distribution of nodal lines and Weyl points}
A movie demonstrates the momentum space distribution of nodal lines calculated without spin-orbit coupling (SOC) and Weyl points with SOC. The movie can be download here: 
\url{http://pan.baidu.com/s/1pJwTsmB}.
\end{document}